\documentstyle[12pt]{article}

\begin{document}

\begin{titlepage}
\begin{flushright}
{\large \bf UCL-IPT-98-09 \\ CINVESTAV-FIS-98-09} 
\end{flushright}
\vskip1.5cm
\begin{center}

{\Large \bf Arithmetic and the standard electroweak theory}
\vskip 2cm

{\large G. L\'opez Castro$^{a,\ }$\footnote{e-mail: {\it
glopez@fis.cinvestav.mx}} and J. Pestieau$^{b,\ }$\footnote{e-mail: {\it
pestieau@fyma.ucl.ac.be}}} \\

$^a$ {\it Departamento de F\'\i sica, Centro de Investigaci\'on y de
Estudios} \\ {\it Avanzados del IPN, Apdo. Postal 14-740, 07000 M\'exico,
D.F., M\'exico}
\

$^b$ {\it Institut de Physique Th\'eorique, Universit\'e Catholique }\\
{\it de Louvain, Chemin du Cyclotron 2,
B-1348, Louvain-la-Neuve, Belgium} \end{center}
\vskip1.5cm
\begin{abstract}
We propose the relations, $1/e - e = 3$ and $\mbox{tan} 2 \theta_W = 3/2$,
where $e$ is the positron charge and $\theta_W$
is the weak angle. Present experimental data support these relations to a
very high accuracy. We suggest that some duality
relates the weak isospin and hypercharge gauge groups of the standard
electroweak theory.

\end{abstract}


\end{titlepage}%

\medskip

\

In the $SU(2) \otimes U(1)$ electroweak theory \cite{1}, two parameters are
necessary to determine the gauge sector. Those parameters are associated 
to the coupling strengths of the weak isospin and hypercharge gauge
groups, and are not related among themselves in the electroweak theory. In
the following, we choose them to be $\alpha$ \cite{2} and $\cos
\theta_W$: 
\begin{equation}
\alpha \equiv \frac{e^2}{4\pi} = [137.03599944(57)]^{-1} \end{equation}
where $e$ is the positron charge, and
\begin{equation}
\cos \theta_W \equiv \frac{m_W}{m_Z} = 0.8820 \pm 0.0009, \end{equation}
with \cite{3}
\begin{eqnarray*}
m_Z &=& 91.1867 \pm 0.0020 \ \mbox{GeV} \\ m_W &=& 80.43 \pm 0.08 \ \mbox{GeV}.
\end{eqnarray*}
In the following we consider simple relations satisfied by the positron
charge $e$ and the weak angle $\theta_W$, which reproduce the experimental
values of these quantities to a high accuracy.

First let us consider the equation
\begin{equation}
y - \frac{1}{y} = 3.
\end{equation}
The two solutions of Eq.(3) are
$$ y_\pm = \frac{3\pm \sqrt{13}}{2} =
\left\{ \begin{array}{c} 3.302775638 \\
-0.302775638,
\end{array} \right.
$$
with $y_+ = -(y_-)^{-1}$, as it is immediately seen by inspection of
Eq.(3). We note that $y_+$ is a continuous
fraction given by $$
y_+ = 3+\frac{1}{3+\frac{1}{3+\frac{1}{3+\frac{1}{3+\cdots}}}}. $$

Let us define a positron charge $\bar e$ and a weak angle $\bar \theta_W$
which we {\it assume} to satisfy the
following relation \begin{equation}
\frac{1-\mbox{tan} \bar \theta_W}{1+\mbox{tan} \bar\theta_W} = \bar e = -y_- =
(y_+)^{-1}= 0.302775638\ .
\end{equation}
$\bar e$ can to be compared with
\begin{equation}
e = (4\pi \alpha)^{1/2} = 0.302822121
\end{equation}
as obtained from Eq.(1). We obtain
$$
\frac{e}{\bar e} = 1.00015\ ,
$$
{\it i.e.}, $\bar e$ reproduces to an impressive accuracy the value of the
positron charge.

From Eq.(4), we can evaluate the weak angle $\bar \theta_W$,
\[
\tan \bar \theta_W = \frac{1-\bar e}{1+\bar e} = 0.535184, \]
or equivalently:
\begin{eqnarray}
\sin \bar \theta_W &=& \frac{1-\bar e}{\sqrt{2 (1+\bar e^2) }} =
0.471858\ , \\
\cos \bar \theta_W &=& \frac{1+\bar e}{\sqrt{2 (1+\bar e^2)}} = 0.881674\ .
\end{eqnarray}
Again, the impressive agreement between Eqs. (7) and (2) provides support
for the assumption contained in Eq. (4).

By combining Eqs. (3) and (4), we can get very simple relations to
determine the absolute values of $\bar e$ and $\bar \theta_W$, namely
\begin{equation}
\frac{1}{\bar e} -\bar e = 3
\end{equation}
and
\begin{equation}
\tan 2\bar \theta_W = \frac{3}{2} \ .
\end{equation}
These two equations summarize our main results.

A relation between the positron and the down quark electric charges can
also be obtained by introducing $\bar e_d = -\bar e/3$ into Eq. (8). We
obtain:
\[
\frac{1}{\bar e_d} = 1 - \Biggl(\frac{1}{\bar e}\Biggr)^2. \]

The quark fractional charge is given in term of the square of electron
charge.

It is worthwhile to consider now the running coupling $\alpha (q^2)$
defined as \cite{4}
\[
\alpha (q^2) = \frac{\alpha}{1 +\frac{\textstyle
\Sigma_{\gamma}(q^2)}{\textstyle q^2} -\Sigma'_{\gamma}(0)}\ . \]
The expressions of the photon self-energy $\Sigma_\gamma (q^2)$, at the
one-loop level, can be found in appendix A of Ref. \cite{4}. At the
electron mass scale
we obtain:
\begin{eqnarray}
\alpha^{-1} (m_e^2) &=& \alpha^{-1} + \frac{17}{9\pi}-\frac{1}{\sqrt{3}}
\nonumber \\
&=& 137.0599011\ .
\end{eqnarray}
Then
\begin{equation}
e (m_e^2) \equiv \Bigl(4\pi \alpha (m_e^2)\Bigr)^{1/2} = 0.302795715,
\end{equation}
can be compared with $\bar e$ given in Eq.(4): $$
\frac{e(m_e^2)}{\bar e} = 1.000066.
$$
The agreement between $e(m_e^2)$ and $\bar e$ is even more impressive than
between $e$ and $\bar e$. Indeed, $\bar e$ coincides with the running
positron charge at the scale $\sqrt{q^2} \approx 1.27 m_e$.

Defining
\begin{equation}
\bar \alpha \equiv \frac{\bar e^2}{4\pi} = [137.0780788]^{-1} \end{equation}
with $\bar e$ given in Eq.(4), we find that \begin{equation}
\alpha = \bar \alpha \left ( 1+\frac{16\bar \alpha}{121 \pi} \right ) =
(137.0360012)^{-1}\ ,
\end{equation}
provides a relation between $\alpha$ and $\bar \alpha$ to an excellent
accuracy.

To conclude, let us comment that Eq.(3) belongs to the class of equations
\begin{equation}
y - \frac{1}{y} = n \ \ \ \ (n \ \mbox{is a real number}). \end{equation}
When $n=1,$ $y_+ = \frac{1+\sqrt{5}}{2}$ is the famous golden number
\cite{5}. For any $n,$ the two solutions of Eq.(14) exhibit the property
$y_+ y_- = -1$ (and, on the other hand, $y_+ + y_- = n$), with $$
y_\pm = \frac{n \pm \sqrt{n^2+4}}{2}.
$$
Notice that $y_+$ can be represented by a simple continuous fraction $$
y_+ = n + \frac{1}{n+\frac{1}{n+\frac{1}{n+\cdots}}} \ . $$
Because Eq.(14) implies
$$
y^2_\pm = ny_\pm + 1\ ,
$$
then
\begin{equation}
y^{m+2}_\pm = n y^{m+1}_\pm + y^m_\pm.
\end{equation}
From Eq.(15), we get by recurrence, the Fibonacci series \cite{5}, when
$n=1$
and $m$ is an integer. With $n=3$, we get $\bar e^{m+2} + 3 \bar e^{m+1} =
\bar e^m$, relating the different terms in the series expansions of Q.E.D.
Note that the dual relation $y_+ y_- = -1$ reminds the relation between the
magnetic charge $g$ of the Dirac monopole \cite{6} and the electric charge
$e : ge =
\frac{\mbox{\footnotesize integer}}{2}$, if the integer is equal to --2.
 The dual property of the solutions to Eq. (3) and Eq. (4) together with
the excellent agreement between Eqs. (8)--(9) and experimental data 
suggest that some duality may be hidden relating the weak isospin and
hypercharge gauge groups of the standard electroweak theory.

\end{document}